\documentclass[%
reprint,  
superscriptaddress,
 amsmath,amssymb,
 aps,
floatfix,
]{revtex4-2}

\usepackage{graphicx}
\usepackage{dcolumn}
\usepackage{bm}
\usepackage{hyperref}
\hypersetup{
	colorlinks=true,
	linkcolor=blue,
	citecolor=blue,
	urlcolor=blue}
\usepackage{xcolor}

\begin{document}

\title{Ion heating and flow driven by an Instability found in Plasma Couette Flow}

\author{J.~ Milhone}
\author{K.~ Flanagan}%
\affiliation{Department of Physics, University of Wisconsin--Madison, 1150 University Avenue, Madison, WI 53706, USA}
\author{J.~ Egedal}
\affiliation{Department of Physics, University of Wisconsin--Madison, 1150 University Avenue, Madison, WI 53706, USA}
\author{D.~ Endrizzi}
\affiliation{Department of Physics, University of Wisconsin--Madison, 1150 University Avenue, Madison, WI 53706, USA}
\author{J.~ Olson}
\affiliation{Department of Physics, University of Wisconsin--Madison, 1150 University Avenue, Madison, WI 53706, USA}
\author{E.~E.~ Peterson}
\affiliation{Plasma Science and Fusion Center, Massachusetts Institute of Technology, 77 Massachusetts Avenue, NW 17 Cambridge, MA 02139}
\author{J.~C.~Wright}
\affiliation{Plasma Science and Fusion Center, Massachusetts Institute of Technology, 77 Massachusetts Avenue, NW 17 Cambridge, MA 02139}
\author{C.~B.~ Forest}
\affiliation{Department of Physics, University of Wisconsin--Madison, 1150 University Avenue, Madison, WI 53706, USA}

\date{\today}

\begin{abstract}
We present the first observation of instability in weakly magnetized, pressure dominated plasma Couette flow firmly in the Hall regime. Strong Hall currents couple to a low frequency electromagnetic mode that is driven by high-$\beta$ ($>1$) pressure profiles. Spectroscopic measurements show heating (factor of 3) of the cold, unmagnetized ions via a resonant Landau damping process. A linear theory of this instability is derived that predicts positive growth rates at finite $\beta$ and shows the stabilizing effect of very large $\beta$, in line with observations.
\end{abstract}

\maketitle


Couette flow has long been studied for its practical importance {\sl and} its elegant simplicity. In Taylor's seminal paper~\cite{Taylor1923}, the stability of Taylor-Couette flow, the differential rotation between two rotating cylinders, is derived, providing a cornerstone of hydrodynamic theory that accurately predicts the onset of dynamic flow structures. Extending beyond traditional fluids, Taylor-Couette flow is studied in a wide variety of mediums ranging from visco-elastic polymers~\cite{Boldyrev2009} to liquid metals~\cite{Nornberg2010,Sisan2004a,Spence2006} where the dynamic properties of these novel materials can be verified. 
The stability of a conducting fluid Couette flow, was analyzed in the magnetohydrodynamic~(MHD) framework both by Chandrasekhar~\cite{Chandrasekhar1960a} and Velikhov~\cite{Velikhov1959a}. In the presence of a weak magnetic field, this flow can become unstable, to the magnetorotational instability (MRI), which is thought to drive turbulent angular momentum transport in accretion disks~\cite{Shakura1973,Balbus1991a}. 

The experimental search for the MRI and other flow-driven MHD instabilities has been primarily led by liquid metal Couette flow experiments~\cite{Donnelly1964,Roach2012a,Liu2006,Stefani2009}. Recently, however, Taylor-Couette flow has been demonstrated in an unmagnetized plasma~\cite{Collins2012, Flanagan2020}. Plasma Couette flow opens up an entirely new set of dynamics that are important for connecting laboratory measurements to flow-dominated astrophysical plasmas. In the diffuse limit where Hall effects become important, ions and electrons decouple at scales below the ion inertial scale ($d_i = c / \omega_{pi}$), and the magnetic field is frozen into the moving electron current rather than the ion fluid. In the case of the Hall MRI, positive growth rates are restricted to configurations where the magnetic field and flow angular momentum are antiparallel~\cite{Wardle1999, Ebrahimi2011, Flanagan2015}. Other instabilities, such as the Simon-Ho~\cite{Simon1963,Hoh1963}, Farley-Buneman~\cite{Farley1963,Buneman1963a}, and gradient drift instability~\cite{Frias2013} can develop in Hall plasmas where large electron drifts act as a source of free energy.

\begin{figure}[t!]
    \centering
    \includegraphics[width=3.375in]{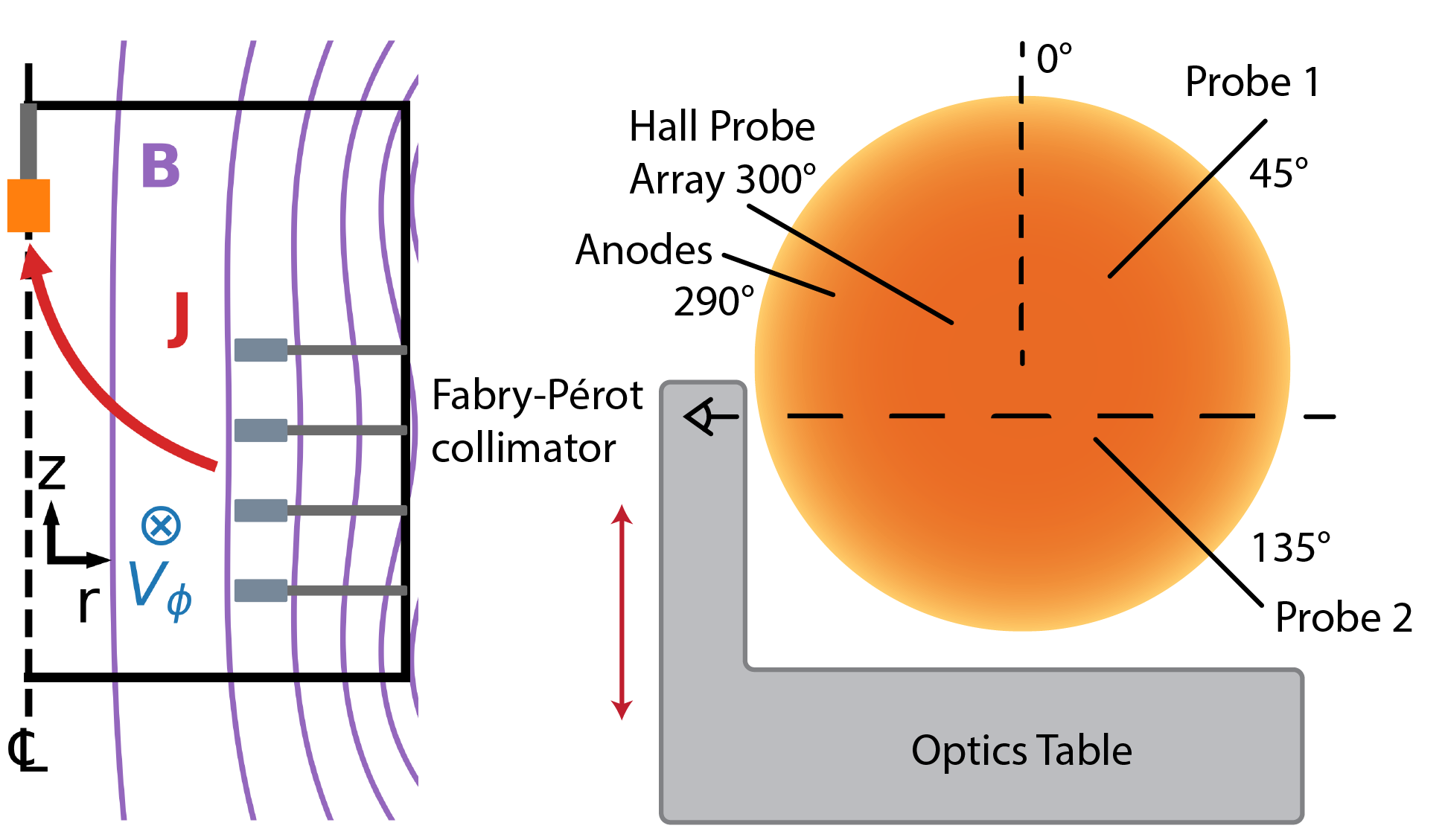}
    \caption{(Left) Volumetric flow drive implemented on PCX. 
    (Right) Top down schematic of PCX displaying azimuthal location of diagnostics. 
    }
    \label{fig:PCX_setup}
\end{figure}

In this Letter, we describe the observation of a new instability in a confined, high-$\beta$ ($\beta$ is the ratio between kinetic and magnetic pressures in a plasma), plasma Couette flow firmly in the collisionless Hall regime. A unique, dynamic equilibrium is created using volumetric flow drive generated by a radial current crossing a weak magnetic field. The direction of radial current has been shown to have large qualitative effects associated with the Hall regime, namely that outwardly directed current creates centrally peaked, stable flows despite the favorable alignment of the magnetic field and flow angular momentum for the Hall MRI~\cite{Flanagan2020}. The present paper is concerned with the opposite current drive direction that results in
a rigid-body ion flow profile and a very strong electron current that expels most of the applied magnetic field. Remarkably, we find this weakly flowing, mostly magnetic-field-free system to be unstable. 
Large fluctuations of the plasma density and axial magnetic field are observed that we attribute to an electromagnetic instability driven by the electron pressure and $E \times B$ drifts. These electron drifts interact via a two-stream-like instability with the cold, unmagnetized ions. Profiles of ion heating and flow are precisely measured by a Fabry-P\'erot spectrometer~\cite{Milhone2019} and document both ion heating and an elongated tail at velocities above the phase velocity of the wave that is consistent with the dissipation of the saturated instability via Landau damping. 

This study was performed on the Plasma Couette Experiment (PCX)~\cite{Collins2012,Collins2014,Flanagan2015} shown in Fig.~\ref{fig:PCX_setup}(a), which operates at the Wisconsin Plasma Physics Laboratory (WiPPL)~\cite{Forest2015a}. A 1~second argon plasma is created with a hot lanthanum hexaboride (LaB$_6$) cathode inserted from the top on axis and biased relative to cold molybdenum anodes placed near the outer plasma edge. Radial current is drawn across an externally applied vertical magnetic field applying a torque to the entire plasma resulting in toroidal flow. While the magnetic field in the main volume of the plasma is weak (typically $<$10~G), an array of permanent magnets on the boundary of the plasma provide bucket-like confinement~\cite{Cooper2014}.

The experiment has an array of magnetic, electrostatic, and optical diagnostics, with toroidal positions shown in Fig.~\ref{fig:PCX_setup}(b). A linear array of 15 3-axis Hall probes~\cite{Peterson2019a} with 1.5 cm spacing and a frequency response from DC up to 100 kHz is mounted at $\phi=300^{\circ}$. Two electrostatic probes mounted at azimuthal angles of $45^{\circ}$ and $135^{\circ}$ and separated by 18.5~cm in $z$ measure ion saturation current to determine plasma density. The Fabry-P\'erot spectrometer is mounted on a scanning optics table for measuring the ion temperature and velocity profiles. This combination of probe locations allows for radial profile measurements at multiple $z$ and $\phi$ locations, helping to identify azimuthal and axial symmetries. 

\begin{figure}
    \centering
    \includegraphics{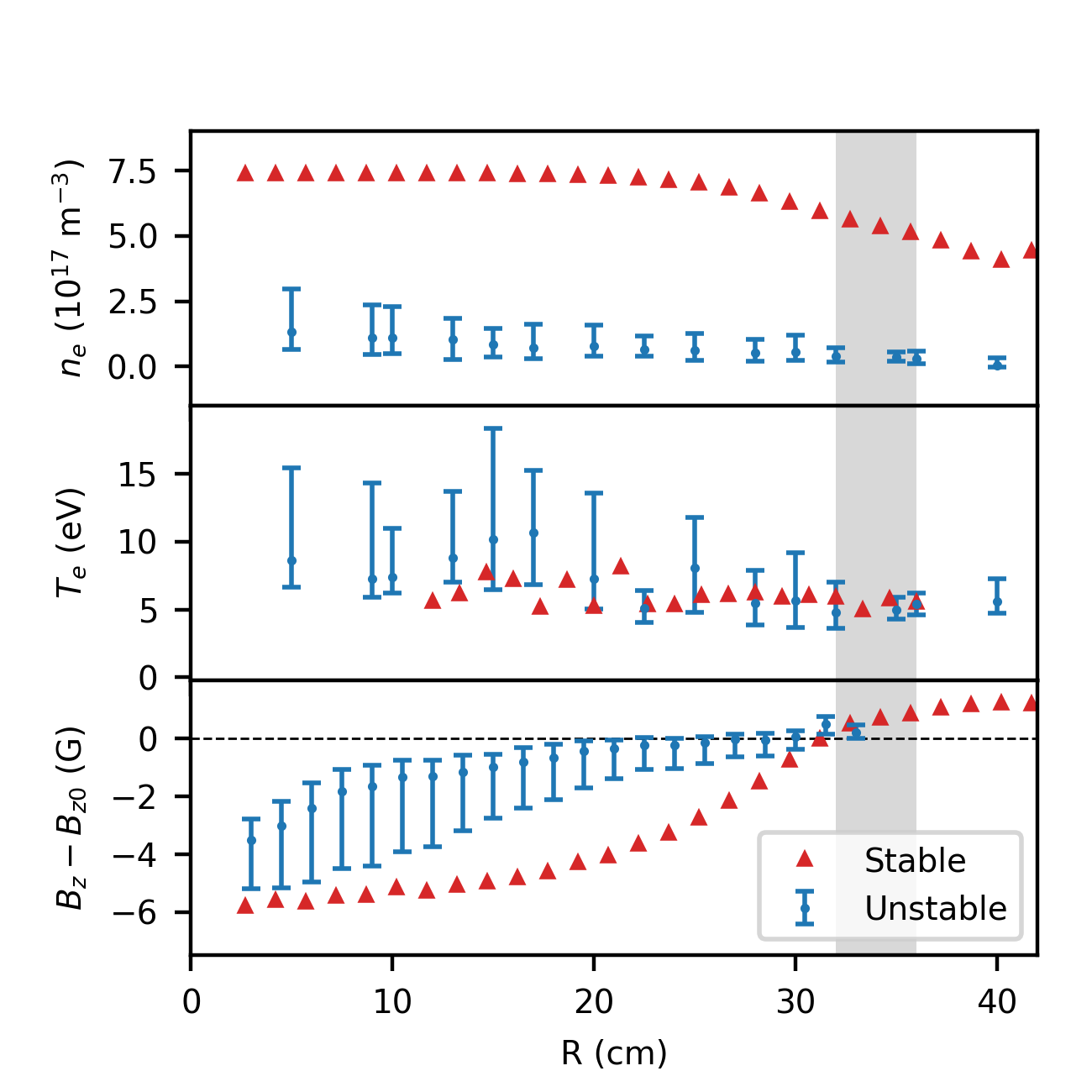}
    \caption{Profiles of $n_e$, $T_e$, and $\Delta B_z$. Error bars indicate the median of the fluctuations with the upper and lower limits indicating the extrema of the fluctuations. The grey shaded region indicates the anode location. The vacuum magnetic field is roughly 6~G for the stable case and 10~G for the unstable case.}
    \label{fig:profiles}
\end{figure}

Our high-$\beta$ equilibrium exists in a recently discovered Hall dominated regime~\cite{Flanagan2020}. The radial electric field used to power the system and drive radial currents has the secondary effect of also generating a Hall current in the azimuthal ($\phi$) direction; this current in turn induces a large magnetic field in opposition to the applied field, completely demagnetizing the core. In the stable case as seen in Fig.~\ref{fig:profiles}, most of the initial magnetic field is removed from the core of the plasma. However, when instability is observed, less magnetic field is removed while large fluctuations are present across the plasma profile implying that the Hall currents are reduced. As part of the equilibrium, the density gradient extends much deeper into the plasma compared to previous multi-cusp confinement experiments~\cite{Collins2014,Cooper2014,Cooper2016a,Weisberg2017}. For the measured ion temperatures, the gyroradius, $r_L=V_{thi}/\Omega_{ci}\sim 40$~cm, and ion inertial scale, $d_i=c/\omega_{pi}\sim~100$~cm, are both on the order of the system size, but the collisional mean-free path is even longer. Hence, the ions are collisionless but unmagnetized. In contrast, the electrons have gyroradii that are less than a cm except for very weak fields ($<$ 0.5 G), and thus, are fully magnetized over much of the profile. As a result, the plasma is firmly in the Hall dominated regime where the magnetic field is frozen into the electron fluid. 

\begin{figure}[htbp]
    \includegraphics[width=3.375in]{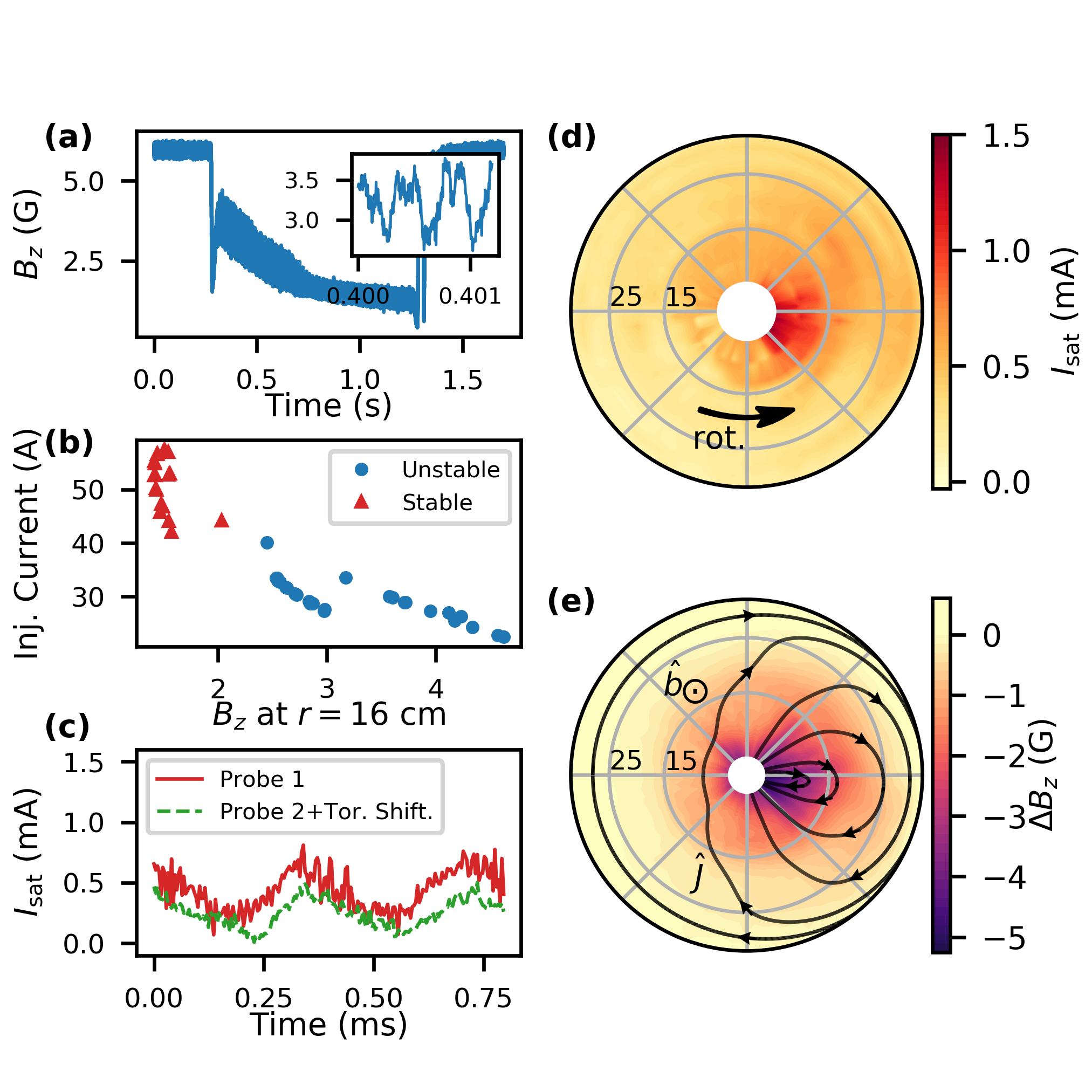}
    \caption{(a) Time trace of $B_z$ at $r=16$ cm.
    (b)Injected current vs magnetic field.
    (c) $I_{\rm{sat}}$ current measurements from two electrostatic probes. 
    (d) 2D reconstruction of the ion saturation current, $I_{\rm{sat}}$.
    (e) 2D reconstruction of the magnetic field, $\Delta B_z$, and current density streamlines.
    }
    \label{fig:massive_figure}
\end{figure}

Large magnetic fluctuations are measured with a frequency ranging from 2-3 kHz, which lies between the ion cyclotron frequency, $\Omega_{ci}$, and the electron cyclotron frequency, $\Omega_{ce}$. The magnetic fluctuations are predominately in $B_z$ with no measured fluctuations in the $r$ and $\phi$ components. A time trace of the fluctuations can be seen in Fig.~\ref{fig:massive_figure}(a). The applied field is reduced from the initial $\sim 6$~G field and has approximately $\sim 0.75$ G fluctuations. Figure~\ref{fig:massive_figure}(b) shows a scaling of injected current for two cases of magnetic field with varying cathode bias voltage. Larger injected currents decrease the magnetic field and lead to stable plasmas while unstable plasmas show a large reduction in injected current and resulting toroidal current. High-speed video imaging~\cite{SupplementalVideo} 
shows a dense structure rotating around the central axis of PCX indicating a predominantly flute-like ($\lambda_z \sim 0$) $m=1$ mode (one period in $\phi$). Measurements of the rotation period match the measured fluctuation frequencies from the magnetic and electrostatic probes. 

We observe that with a positive applied magnetic field (in the $+\hat{z}$ direction), the mode rotates in the anti-clockwise direction when viewed from above. Ion saturation current fluctuations are measured by the two electrostatic probes separated by $90^{\circ}$ azimuthally and $18.5$ cm vertically.  As shown in Fig.~\ref{fig:massive_figure}(c), the probes show good agreement with an $m=1$ flute mode with $k_z =2 \pi/\lambda_z \sim 0$ when the azimuthal lag is accounted for.

Reconstructions of the evolved instability reveal a large amplitude ($\tilde{n}/n \sim \tilde{b}/B\sim 1$) rotating structure. Figures~\ref{fig:massive_figure}(d, e) present a 2D reconstruction of the mode structure. Measurements of density and magnetic field were made by scanning probes radially, and each time trace was filtered at the mode frequency determined by the fast Fourier transform of $B_z$ and synced by maximizing the cross correlation in $\delta B_z$. The mode structure was then recovered by mapping time to the azimuthal direction ($2 \pi f t \rightarrow \phi$). The mode shown in Figs.~\ref{fig:massive_figure}(d, e) includes the toroidal lag between the Hall probe array and Probe 2. The peak density of the mode has a much steeper gradient compared to the stable equilibrium and is out of phase with the magnetic field where most of the magnetic flux is pushed out.

The non-linear saturated state can be understood as a rotating structure well defined by a force balance between the $J\times B$ force from the currents associated with the magnetic field depression, the electron pressure gradient, and an electric field that points in the same direction as the density gradient. 
This saturated state is consistent with an electron vortex being advected hydrodynamically, with parallels to Helmholtz's theorem for vorticity advection.  

\begin{figure}[t!]
    \includegraphics[width=3in]{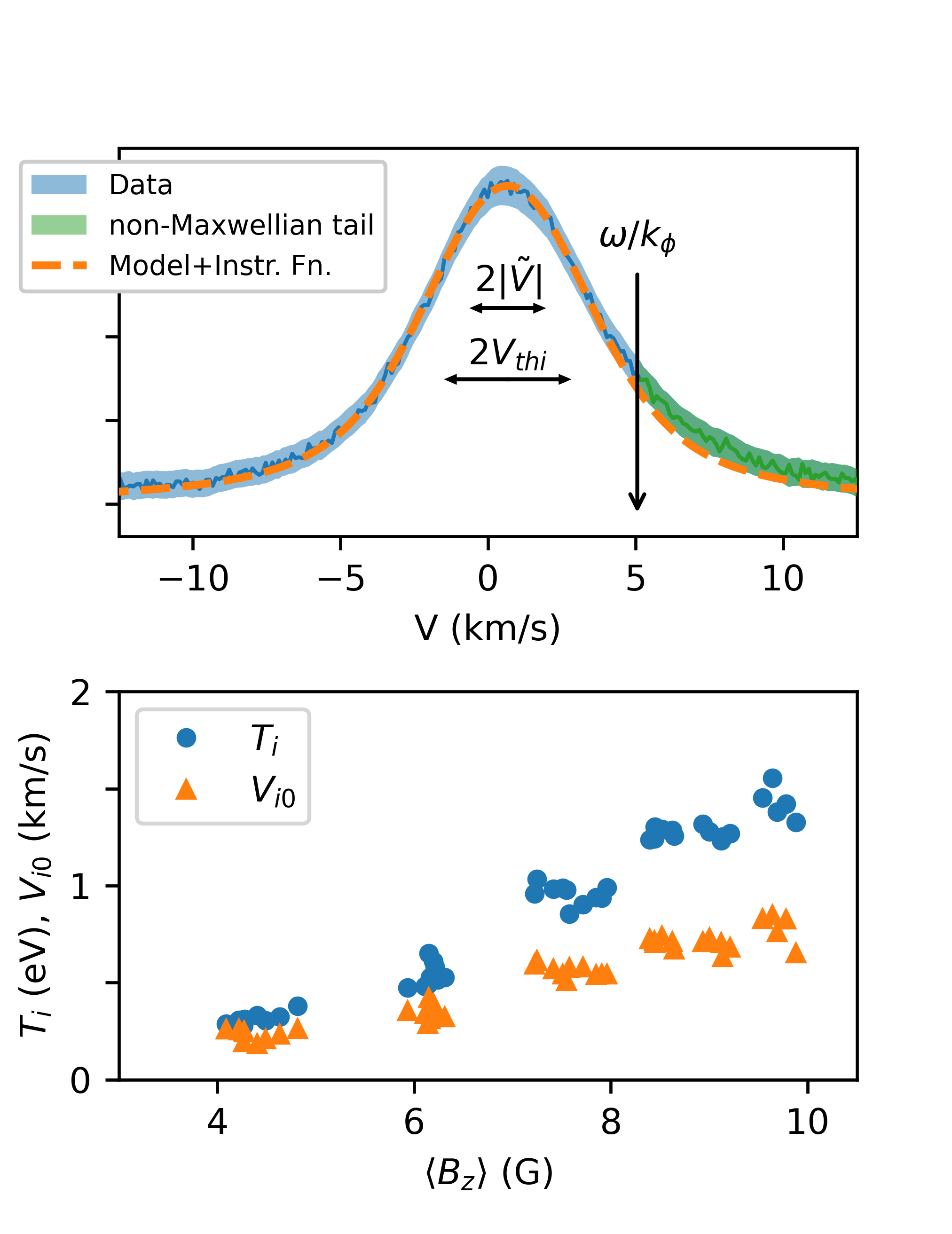}
    \caption{(top) Spectrum from the Fabry-P\'erot spectrometer. Shaded regions represent the total error in the measurement. 
    (bottom) $T_i$ and peak $V_{i0}$ vs the mean magnetic field at $r=14$ cm. 
    }
    \label{fig:fp summary}
\end{figure}

Measurements of the ion temperature and velocity were made with a high-resolution, large \'etendue Fabry-P\'erot spectrometer 
(for a detailed explanation of this diagnostic and its capabilities, see~\cite{Milhone2019}). Spectra of Doppler shifted emission from ArII ions were taken with integration times much longer than the period of oscillation at various tangency radii. A measured spectrum mapped to velocity space is shown in the top of Fig.~\ref{fig:fp summary}. The spectrum shown is integrated along a chord with a tangency radius of 30~cm near the edge of the plasma. The sharp gradient in the edge of the plasma means that the emissivity along this chord is only non-negligible over a small region, allowing us to approximate this as a single point measurement at a radius of 30~cm. Extensive modeling of any profile effects near the edge confirm that this approximation is within the measurement error of the spectrometer. The spectrum is modeled as a thermally broadened, drifting Maxwellian distribution with an oscillating quiver velocity, $\tilde{V}$. Peak velocities of $\sim 1$~km/s are measured with deviations at velocities on the order of a few $V_{th,i}$ in the distribution function near the phase velocity of the wave. The measured ion temperature is $1.4$~eV which is 2-4x higher than a typical stable PCX argon plasma with similar electron density and temperature. We also note the significant deviation of the spectrum above the phase velocity (see the green shaded data in the top of Fig.\ \ref{fig:fp summary}) of the wave which we attribute to quasi-linear flattening of the distribution function. This deviation is not present on any spectra collected from non-fluctuating equilibria. As presented in bottom plot of Fig.~\ref{fig:fp summary}, the ion temperature and velocity vary linearly with the mean magnetic field in the plasma. 

The onset of these fluctuations is consistent with a new instability, which we name the electromagnetic gradient drift instability~(EGDI). The instability is driven by magnetized electrons drifting through a background of unmagnetized ions and is similar in some ways to the low-$\beta$ electrostatic electron drift instability derived in ~\cite{Frias2013,Frias2013a}. When $\beta$ is high, gradients in the equilibrium magnetic field strength contribute significantly to enhancing electron drift velocities, and the instability can develop strong fluctuations in $B_z$ with finite ${\bf k_{\perp}}$. The magnetic fluctuations are related to the density and electrostatic potential fluctuations through Amp\`ere's law assuming the radial current is purely from the fluctuating electron drift velocities and neglecting the ultra-high frequency displacement current. We have derived a local dispersion relation detailed in~\cite{SupplementalDerivation}, that shows the instability is driven by magnetic field and density gradients, parametrized by inverse gradient length scales, $\kappa_B=\partial \ln{B_{0}}/\partial r$ and $\kappa_n=\partial \ln{n}/\partial r$, respectively. 

\begin{figure}
    \centering
    \includegraphics{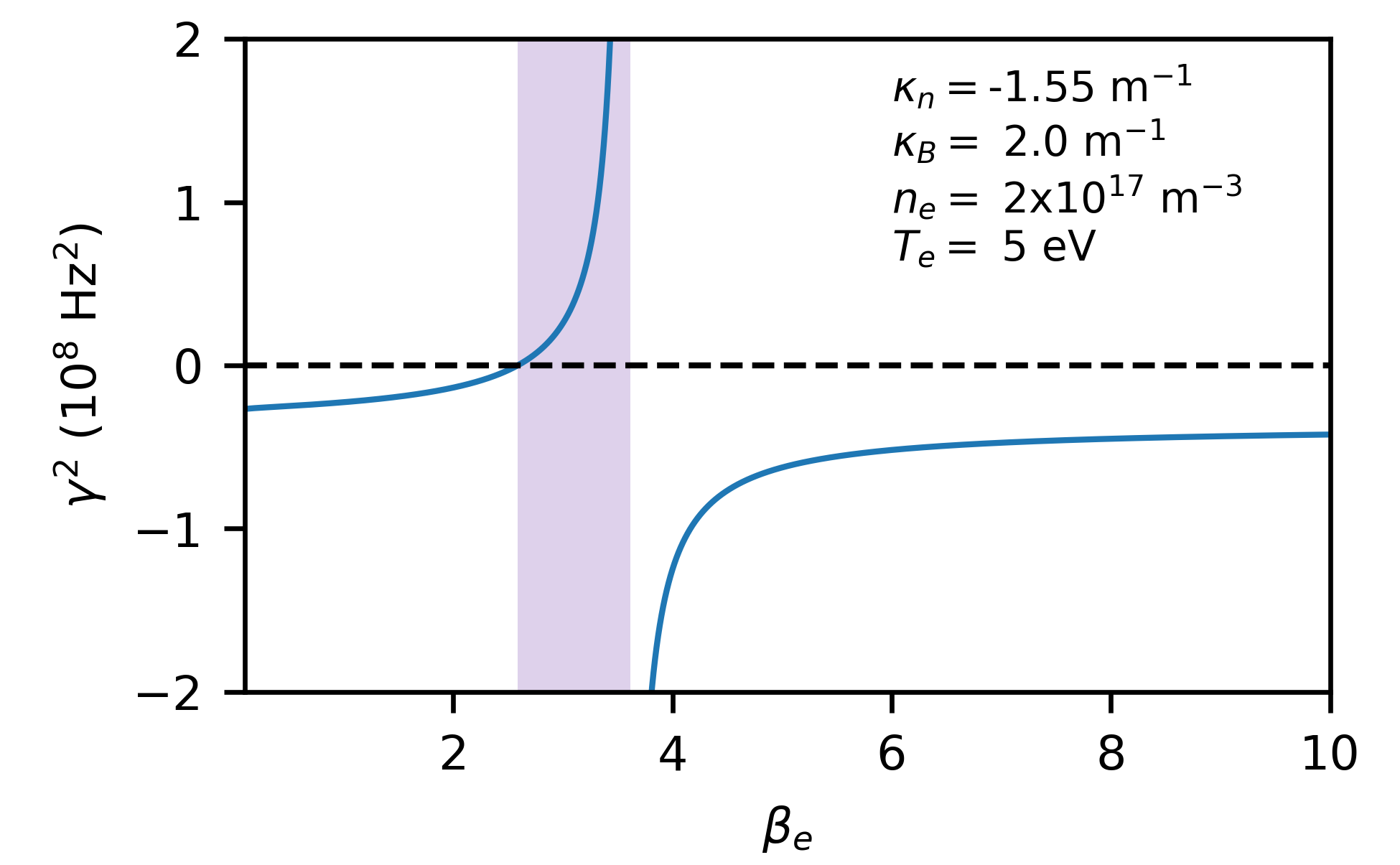}
    \caption{Linear growth rate squared, $\gamma^2$, of the Electromagnetic Gradient Drift Instability, vs $\beta$; Unstable if~$\rm{Re}(\gamma) > 0$. The shaded region is the range of $\beta$ where the instability is expected.}
    \label{fig:growth_rate_vs_beta}
\end{figure}

The linear growth rate of the mode is heavily dependent on the magnitude of $\beta$. For $\beta \ll 1$, PCX is completely stable with $\kappa_B >0$ and $\kappa_n <0$ as seen in Frias et al~\cite{Frias2013}. For moderate $\beta$, the plasma is unstable to this electromagnetic mode, however, at large enough $\beta$, it is stabilized by the large plasma pressure, consistent with experimental results shown in Fig.~\ref{fig:massive_figure}(b). For realistic parameters on PCX, $\kappa_n=-1.55$~m$^{-1}$, $\kappa_B=2$~m$^{-1}$, $n_e = 2 \times 10^{17}$~m${}^{-3}$, the plasma is unstable for a small range of $\beta$, and the growth rate diverges at the high-$\beta$ stability boundary as shown in Fig.~\ref{fig:growth_rate_vs_beta}.


The observed ion heating and flow are consistent with energy and momentum transport from the wave into the ions. Normally, hot filament plasmas are very cold ($\ll 1$~eV) since the collisional heating from the electrons ($\sim 2$~W for these plasmas) is weak and heat losses via collisions with neutrals and charge-exchange are relatively large. Since the wavelength of the mode is on the order of the mean free path, the use of Braginskii's unmagnetized viscosity~\cite{Braginskii1965} for collisional viscous heating is invalid indicating a need for a collisionless process such as Landau damping. The heating from Landau damping can be estimated from quasi-linear diffusion \cite{Bernstein1966} given the measured strength of the oscillating electric field (measured in these plasmas to be on the order of 10 V/m). For a single wavenumber, $k$, the power transferred from the oscillating field to the ions is
\begin{equation}
     P = \mathbb{V}\sqrt{\frac{\pi}{2}}\,\omega_{pi}^{2}\frac{\epsilon_{0}|E|^{2}}{ kv_{thi}}\frac{v_{ph}^{2}}{v_{thi}^{2}}\,e^{-\frac{1}{2}\frac{v_{ph}^{2}}{v_{thi}^{2}}}\quad,
     \label{eq:landau_pow}
\end{equation}
where $\mathbb{V}$ is the volume, $v_{thi}\equiv\sqrt{kT_{i}/m_{i}}$ is the ion thermal speed, $\omega_{pi}\equiv\sqrt{ne^{2}/\epsilon_{0}m_{i}}$ is the ion plasma frequency and $v_{ph}\equiv\omega/k$ is the phase velocity of the wave. Choosing a wavenumber of $k\simeq1$~m$^{-1}$, which corresponds to a $k_{\phi}$ for the full circumference at roughly 15~cm, the estimated power from Eq.~\ref{eq:landau_pow} is an order of magnitude higher than the collisional heating calculations, approximately 20~W. This Landau heating is more than enough to change $T_{i}$ by the measured amount given simple power balance against neutral collisions and confinement losses. Moreover, the deviation in the distribution function in the direction of mode's phase velocity, as indicated by the shaded green data in Fig.\ \ref{fig:fp summary} (top), suggests the tail is gaining not only energy from the wave but also momentum to cause the ions to rotate with the instability.

In summary, we have observed a unique instability in high-$\beta$, collisionless Hall plasma in the Couette geometry that transfers energy and momentum from the electrons to the ions. This experiment was initially intended to study the magneto-rotational instability where strong flow shear is required as a source of free energy which has not yet been observed. Indeed, for the configuration presented in this paper, the Hall regime MRI is not even expected since the flow angular momentum and the magnetic field~\cite{Balbus2001, Wardle1999c, Ebrahimi2011, Flanagan2015} are not aligned. In the present study, however, we have shown a strong, global instability that is driven by gradients in both the magnetic field and plasma density, not the ion flow.  A local dispersion relationship predicts stability boundaries, but the non-linear structure is clearly global and appears to be a rotating dynamic equilibrium that would benefit from additional modeling.  This theory suggest that strong gradients in the electron drifts effectively take the role of more traditional ion flow shear via a two-stream like instability. With the aid of high-resolution Fabry-P\'erot spectroscopy, we have also shown that this instability transfers energy and momentum to the otherwise cold ion population. This unique Couette instability highlights the importance of the electron fluid on electromagnetic instabilities in the collisionless Hall regime. 

\begin{acknowledgements}
This work was funded in part by the NSF under grant No. 1518115. In addition, this work is supported through the WiPPL User Facility under DOE fund DE-SC0018266.
\end{acknowledgements}

\bibliographystyle{apsrev4-1}


%


\end{document}